\documentclass[runningheads]{llncs}

\usepackage[table,xcdraw]{xcolor}
\usepackage[T1]{fontenc}
\usepackage{amsmath}
\usepackage{enumitem}
\usepackage{amsfonts}
\usepackage{caption}
\usepackage{multirow}
\usepackage{subfig}
\usepackage{hyperref}
\usepackage[squaren,Gray]{SIunits}
\def\doi#1{\href{https://doi.org/\detokenize{#1}}{\url{https://doi.org/\detokenize{#1}}}}

\usepackage{graphicx}
\usepackage{listings}
\newlist{steps}{enumerate}{1}
\setlist[steps, 1]{wide=0pt, leftmargin=\parindent, label=Step \arabic*:, font=\bfseries}

\begin{document}
\title{Anisotropic Hybrid Networks for liver tumor segmentation with uncertainty quantification}
\titlerunning{Anisotropic Hybrid Networks for liver tumor segmentation}

\author{Benjamin Lambert \inst{1, 2} \and Pauline Roca \inst{2} \and 
Florence Forbes \inst{3} \and Senan Doyle \inst{2} \and Michel Dojat \inst{1}}
\authorrunning{B. Lambert et al.}

\institute{Univ. Grenoble Alpes, Inserm, U1216, Grenoble Institut Neurosciences, 38000, FR \and
Pixyl, Research and Development Laboratory, 38000 Grenoble, FR \and
Univ. Grenoble Alpes, Inria, CNRS, Grenoble INP, LJK, 38000 Grenoble, FR}


%
\maketitle              
\begin{abstract} 

The burden of liver tumors is important, ranking as the fourth leading cause of cancer mortality. In case of hepatocellular carcinoma (HCC), the delineation of liver and tumor on contrast-enhanced magnetic resonance imaging (CE-MRI) is performed to guide the treatment strategy. As this task is time-consuming, needs high expertise and could be subject to inter-observer variability there is a strong need for automatic tools. However, challenges arise from the lack of available training data, as well as the high variability in terms of image resolution and MRI sequence. In this work we propose to compare two different pipelines based on anisotropic models to obtain the segmentation of the liver and tumors. The first pipeline corresponds to a baseline multi-class model that performs the simultaneous segmentation of the liver and tumor classes. In the second approach, we train two distinct binary models, one segmenting the liver only and the other the tumors. Our results show that both pipelines exhibit different strengths and weaknesses. Moreover we propose an uncertainty quantification strategy allowing the identification of potential false positive tumor lesions. Both solutions were submitted to the MICCAI 2023 Atlas challenge regarding liver and tumor segmentation. 

\end{abstract}

\section{Introduction}
Liver cancer is the sixth most common cancer worldwide, the fourth most common cause of cancer-related death and hepatocellular carcinoma (HCC) in particular is the most common type of primary liver cancer in adults.
HCC is a cancer with large therapeutic choices depending on the tumor staging. In case of unresectable HCC, transarterial radioembolisation (TARE) is one of the recommended treatments: it consists of the delivering of 20–60 uM-sized yttrium-90 (Y90) microspheres to the tumor arterial supply, causing tumor necrosis through radiation-induced DNA damage and eventual cell death \cite{reig_bclc_2022}. To calculate dosimetry and plan the intervention, the spatial extent of the tumor needs to be determined \cite{smits2015radioembolization}. In clinical routine, the liver and tumor are delineated using contrast-enhanced magnetic resonance imaging with four phases (precontrast, arterial, portal venous, delayed phases) allowing to differenciate the tumor from the liver parenchyma. These tasks are time-consuming and required high medical expertise. 

In this context, automatic tools could reduce intra- and inter-observer variability and help radiologists to save time. Several studies \cite{bousabarah_automated_2021,zheng2022automatic} proposed CE-MRI based deep learning methods to segment liver and HCC tumors using private datasets (of 174 and 190 patients respectively). However, open-source MRI datasets are scarce, and existing ones \cite{clark2013cancer} do not contain manual annotations, inhibiting the development of automated algorithms. In this context, the "A Tumor and Liver Automatic Segmentation" (ATLAS) challenge and dataset have been proposed. The aim of the challenge is to produce automatic segmentation maps of liver and the tumor(s) using CE-MRI of patients with unresectable HCC based on a small annotated training dataset (60 patients) \cite{quinton2023tumour}. This challenge is organised in collaboration with the Medical Image Computing and Computer Assisted Intervention (MICCAI) Workshop on Resource-Efficient Medical Image Analysis (REMIA) which will take place during the MICCAI 2023 edition in Vancouver, Canada. This paper describes our contribution to the challenge, including two different pipelines to segment liver and HCC tumors in CE-MRI. The first pipeline correspond to a multiclass model that simultaneously segment the liver and tumors. In the second pipeline, we used two different binary models, one focusing on the liver and the other on the tumors. Additionally, we propose to complement the segmentations with the estimation of an uncertainty score for each identified tumor lesion, potentially guiding the reviewing of the automated result by the user. 

\section{Materials and Methods}
\subsection{Dataset}

\begin{figure}[b!]
    \centering
    \includegraphics[width=0.8\textwidth]{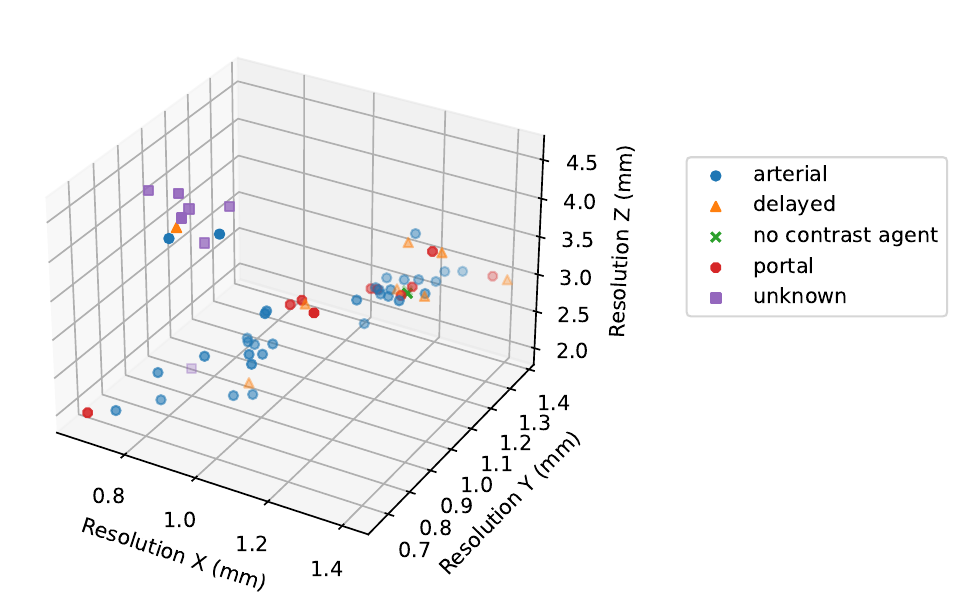}
    \caption{Distribution of the spatial resolution in the training dataset. Colors indicate the corresponding CE-MRI phase. To facilitate visualization, the plot is generated using jitter.}
    \label{fig:resolution}
\end{figure}

The ATLAS challenge dataset is composed of an open-source training fold (60 subjects), allowing the development of segmentation models, as well as an hidden test set (30 subjects) for evaluation \cite{quinton2023tumour}. For each subject, a unique CE-MRI sequence is available, with one of the following phase: arterial, delayed, post-contrast, non-contrast or unknown. A ground truth segmentation mask is associated to each image, containing 3-classes: background, healthy liver and tumor lesions. Training images exhibit variable resolution, ranging from $0.684mm$ to $1.4mm$ in the XY plane, and from $2mm$ to $4.6mm$ in the Z-axis (see Figure \ref{fig:resolution}). To develop our approach, we adopt a 5-fold cross validation scheme. In each fold, 40 subjects are used for training, 8 for validation and 12 for testing. The distribution of data in each fold is resumed in Table \ref{fold_distrib}.

\setlength{\tabcolsep}{1.9pt}
\renewcommand{\arraystretch}{1.2}
\begin{table}[h!]
\centering
\scriptsize
\begin{tabular}{c|ccccc|ccccc|ccccc}
\multicolumn{1}{l|}{} & \multicolumn{5}{c|}{Training (N=40)} & \multicolumn{5}{c|}{Validation (N=8)} & \multicolumn{5}{c}{Test  (N=12)} \\ \hline
Fold & A & D & NC & P & U & A & D & NC & P & U & A & D & NC & P & U \\ \hline
1 & 23 & 6 & 1 & 5 & 5 & 3 & 0 & 1 & 3 & 1 & 7 & 2 & 0 & 2 & 1 \\
2 & 24 & 6 & 1 & 4 & 5 & 2 & 1 & 0 & 4 & 1 & 7 & 1 & 1 & 2 & 1 \\
3 & 24 & 5 & 1 & 4 & 6 & 2 & 1 & 1 & 4 & 0 & 7 & 2 & 0 & 2 & 1 \\
4 & 25 & 5 & 1 & 5 & 4 & 2 & 1 & 1 & 3 & 1 & 6 & 2 & 0 & 2 & 2 \\
5 & 24 & 6 & 0 & 6 & 4 & 3 & 1 & 1 & 2 & 1 & 6 & 1 & 1 & 2 & 2
\end{tabular}
\caption{Data distribution in each fold. A: arterial, D: delayed, NC: non-contrast, P: portal, U: unkwon}\label{fold_distrib}
\end{table}

\subsection{Segmentation pipelines}
\begin{figure}
    \centering
    \includegraphics[width=\textwidth]{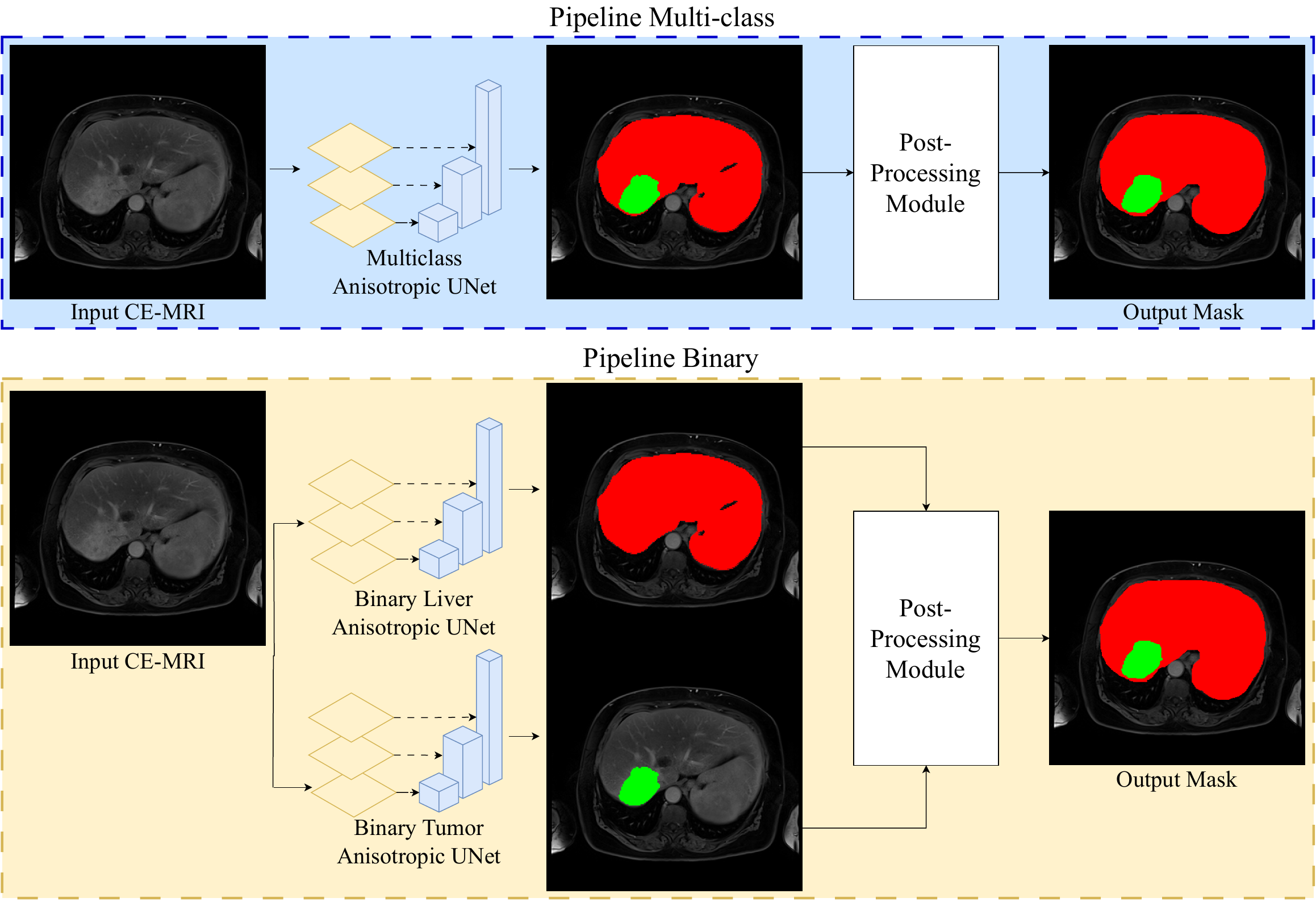}
    \caption{Illustration of our two proposed pipelines, namely multiclass and binary.}
    \label{fig:pipelines}
\end{figure}

In this section, we start by describing the design choices that are common to both pipelines, namely the neural network architecture and loss function. Then, we detail the particularity of both approaches (see Figure \ref{fig:pipelines} for an illustration). 

\subsubsection{Joint design choices}
The first challenge to tackle is the anisotropy of the data, with the Z resolution being up to 4 times that in the plane. Instead of preprocessing the data to obtain a uniform spatial resolution, we instead chose to use a segmentation neural network able to handle the intrinsic anisotropy of the raw data. To this end, our two pipelines revolve around the use of the Anistropic Hybrid U-Net (AHUNet) \cite{liu20183d}. This architecture is composed of a pretrained 2D convolutional encoder ignoring between-slice information, and a 3D convolutional decoder that reconstitute the 3D context. Second, although not the scope of the ATLAS challenge, we aim at enhancing the segmentation of the models by an uncertainty quantification module allowing to estimate a confidence score for each tumor lesion. Thus, the calibration of the output probability is of primary importance, which is usually poor with modern networks \cite{guo2017calibration}. To circumvent this limitation, we resort to using the recently proposed Margin-based Label Smoothing loss \cite{murugesan2023calibrating} which has been proposed to improve the calibration of medical image segmentation models.
 
\subsection{Pipeline Multi-class}
The most direct approach to obtain the segmentation of the liver and tumors is to train a multi-class AHUnet using the 3-classes ground truth masks. We adopt a patch-based approach, following which the input 3D MRI is divided into patches that are processed independently by the model. To guide the selection of the patches size, we computed bounding-boxes around each liver and tumor lesion in the training dataset using the ground truth masks (see Figure \ref{fig:bounding_boxes}). This study indicates that defining a patch size of $256\times256\times64$ allows the encompassment of all tumor lesions, while also encompassing most of the livers. 

\subsection{Dual binary pipeline}
\begin{figure}
    \centering
    \includegraphics[width=\textwidth]{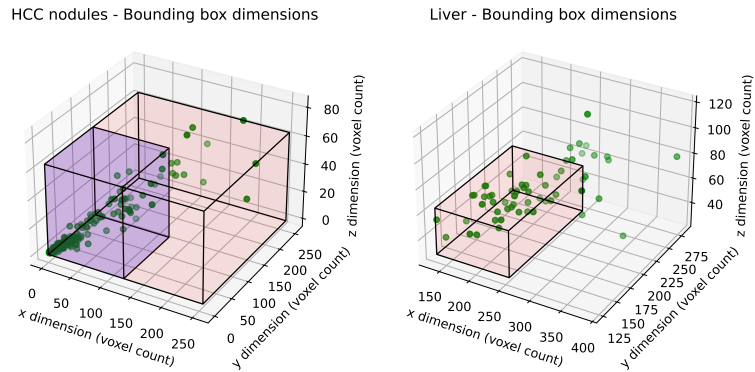}
    \caption{Dimensions of bounding boxes encompassing the tumor lesions (left) and liver (right) on the Atlas dataset. In overlay we present the selected patch sizes of $256\times256\times64$ (red) and $128\times128\times64$  (blue).}
    \label{fig:bounding_boxes}
\end{figure}

Motivated by the observation that the segmentation of the liver was a much more easier task than tumor segmentation, we propose an alternative pipeline that disentangles both tasks. To do so, two distinct binary models are trained: i) a model that segments the \emph{overall} liver (concatenation of the healthy liver class and the tumors) and ii) a model that segments the tumor lesions. From the bounding-box analysis (Figure \ref{fig:bounding_boxes}), we kept the patch size of $256\times256\times64$ for the binary liver model. However, for the binary tumor model, we use a smaller patch size of $128\times128\times64$ as it allows to reduce the imbalance between background and tumor voxels, while still encompassing most of the nodules. At inference, the input MRI is processed by each model. The two resulting binary masks are then aggregated to reconstitute the final 3-class segmentation.

\subsection{Post-processing Module}
Post-processing has been proved as an important step for the automated segmentation of the liver in CT and MRI \cite{furtado2021loss}. We adopt a 3-steps post-processing procedure, which is similar for both pipelines:
\begin{steps}
    \item Only the largest connected component for the liver is kept. A binary morphology \emph{fill holes} operation \cite{2020SciPy-NMeth} is performed on the resulting binary mask.
    \item Tumor lesions outside the liver are removed. This process differs slightly depending on the pipeline. For the multi-class approach, the binary mask of each identified tumor lesion is dilated. If this dilated mask has no intersection with liver voxels, the lesion is deleted. For the binary-approach, the process is more straight-forward, as we simply multiply the liver mask by the tumor mask. As an effect, tumor voxels outside the liver are filtered out.
    \item A binary closing morphological operation is applied to the remaining tumor lesions to improve the smoothness of the borders.
\end{steps}

\subsection{Lesion Uncertainty Quantification}
Additionally to the segmentation of the liver and tumors, we propose to add a lesion uncertainty quantification module to our pipelines. This module complements the segmentation with an uncertainty score for each unique nodule. To achieve this, we collect the tumor probabilities $p_{i,tumor}$ ($i\in[1,N]$) for each of the N voxels of the lesion. The \emph{lesion-wise} uncertainty score is then taken as:

\begin{equation}
    L_{unc} = 1 - \frac{1}{N} \sum_{i=1}^{N} p_{i,tumor}
\end{equation}

\subsection{Evaluation Protocol}
The ATLAS challenge proposes to compare participants based on 9 metrics. The Dice, Surface Dice (SD), Haussdorf Distance (HD) and the Average Symmetric Surface Distance (ASD) are computed for both the \emph{overall} liver (concatenation of the healthy liver and tumor voxels) and the tumors. Lastly, the  Root Mean Square Error (RMSE) on tumor burden is computed, based on the computed volumes for the liver and tumors. To compare each pipeline, we train models on each of the 5 folds, and metrics are aggregated over the 5 test folds.
Finally, qualitative assessment of the computed tumor uncertainty scores is assessed by monitoring the distribution of uncertainties for TP (non-zero intersection with the ground truth tumor mask) and FP (null intersection) lesions. 

\subsection{Ablation Study}
To demonstrate the positive impact of using AHUnet over standard 3D segmentation architecture, we reproduced the multi-class pipeline by replacing the AHUnet by a Dynamic Unet 3D (DynUnet) which has been successful in various medical image segmentation challenges \cite{isensee2018nnu}. We also run the evaluation procedure without post-processing to demonstrate its gain on performance.

\subsection{Implementation Details}
Our framework is implemented in Pytorch \cite{paszke2019pytorch}  and experiments are carried on a Nvidia RTX A5000 GPU. AHUnet and DynUnet are implemented using the open-source MONAI library \cite{cardoso2022monai}. Models are trained using the ADAM optimizer \cite{kingma2014adam} with a learning rate of $2e-4$ and a batch size of $2$, until the loss on validation data stagnates over 30 epochs. During training we use Data Augmentation implemented using the TorchIO library \cite{perez2021torchio} comprising spatial (rotation, flips, affine transforms) as well as intensity (contrast, gamma, bias, noise) augmentations.

\section{Results and Discussion}

\begin{table}[]
\caption{Performance on the 60 subjects aggregated over the 5 test folds. Mean results are presented along the standard deviation.}\label{table:perf}
\centering
\begin{tabular}{c|l|cc|cc||c}
  & & \multicolumn{2}{c|}{Multi-class AHUnet} & \multicolumn{2}{c||}{Dual binary AHUnet} & \multicolumn{1}{c}{DynUnet} \\
  & Post-processing & \checkmark & $\times$ & \checkmark & $\times$ & \checkmark \\ \hline
\parbox[t]{2mm}{\multirow{4}{*}{\rotatebox[origin=c]{90}{Liver}}}  & ASD $\downarrow$ & $2.06\pm1.78$ & $3.31\pm3.53$ & \textbf{1.99$\pm$2.35} & $5.03\pm 5.01$ & $2.57\pm2.70$ \\
& Dice $\uparrow$ & \textbf{0.94$\pm$0.04} & $0.93\pm0.04$ & \textbf{0.94$\pm$0.06} & $0.92\pm0.07$ & $0.93\pm0.06$ \\
&  HD ($\times 10^1$) $\downarrow$& $3.22 \pm1.78$ & $11.27\pm6.03$ & \textbf{2.85$\pm$1.69} & $12.79\pm6.76$ & $3.60\pm2.60$ \\
& SD $\uparrow$& $0.82\pm0.10$ & $0.80\pm0.11$ & \textbf{0.84$\pm$0.12} & 
$0.81\pm0.12$ & $0.81\pm0.13$ \\ \hline
\parbox[t]{2mm}{\multirow{5}{*}{\rotatebox[origin=c]{90}{Tumor}}} & ASD ($\times10^1$) $\downarrow$ & \textbf{1.46$\pm$2.5} & $1.62\pm2.50$  & 1.53$\pm$2.31 & $3.51\pm4.21$ & $1.90\pm2.25$ \\
 &  Dice $\uparrow$ & \textbf{0.57$\pm$0.28} & \textbf{0.57$\pm$0.28} & $0.54\pm 0.29$ & $0.49\pm0.30$ & $0.49\pm0.29$ \\
 & HD ($\times 10^1$) $\downarrow$& \textbf{6.73$\pm$3.84} & $8.87\pm5.74$ & $7.77\pm4.13$ & $18.0\pm7.89$ & $8.49\pm4.34$\\
 &  SD $\uparrow$& 0.40$\pm$0.21 & 0.40$\pm$0.21 & \textbf{0.41$\pm$0.22} & $33.9\pm0.21$ & $0.33\pm0.20$ \\
 & RMSE ($\times 10^{-2}$) $\downarrow$& $ 0.73\pm 1.38$ & $0.75\pm0.01$ & \textbf{0.72$\pm$1.30} & $1.17\pm2.10$ & $1.04\pm2.19$ \\
\end{tabular}
\end{table}

\begin{figure}[htb!]
    \centering
    \includegraphics[width=\textwidth]{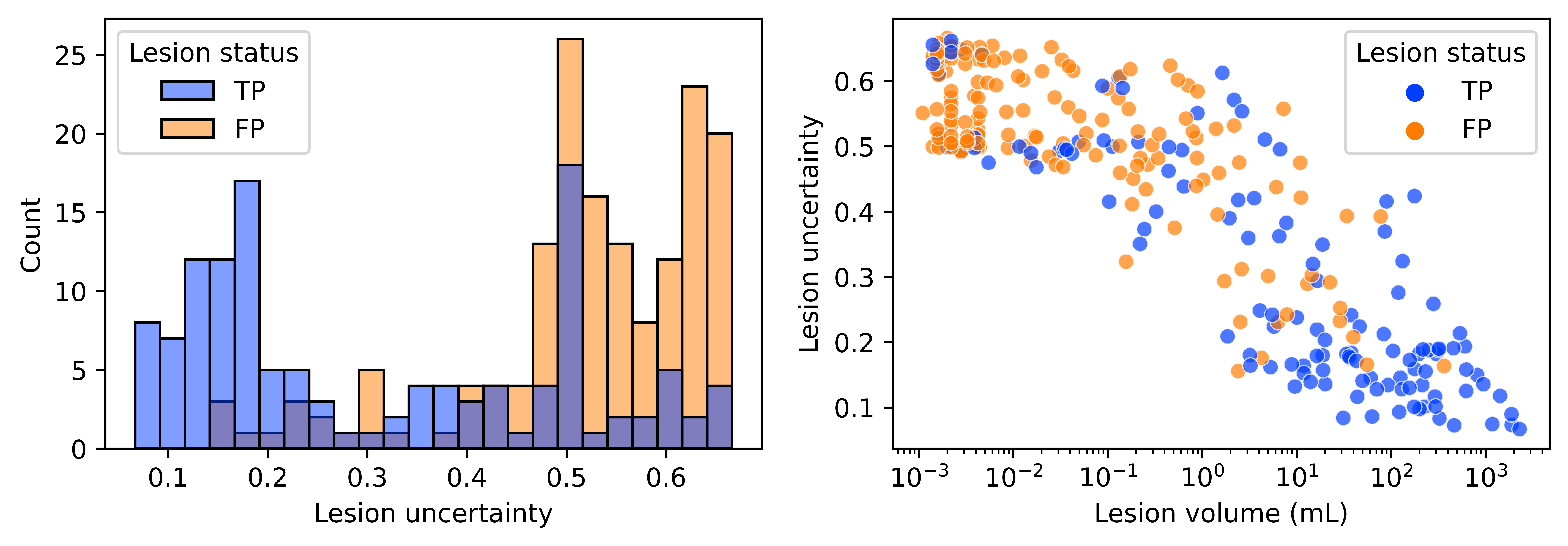}
    \caption{Qualitative evaluation of the tumor lesion uncertainty obtained with the multi-class pipeline. Left: histogram of  uncertainty estimates with respect to the lesion status (TP or FP). Right: lesion uncertainty with respect to the lesion volume (in log-scale).}
    \label{fig:uncertainty}
\end{figure}

Segmentation performance of each pipeline is presented in Table \ref{table:perf}. Qualitative assessment of lesion uncertainty is presented in Figure \ref{fig:uncertainty}. 

For both pipelines, the liver is far better segmented than tumors. On the multi-class AHUnet pipeline for example a mean Dice score of 0.94 is reached for liver segmentation whereas for tumor we reached a value of 0.57.  This can be explained by the small volume of tumors as compared to the liver, and by the important variability in appearance depending on the contrast phase. Moreover, these results are similar to Bousabarah study \cite{bousabarah_automated_2021} that obtained Dice scores per case of 0.91 and 0.48 for liver and tumor segmentation respectively using a deep-learning method based on the four phases of CE-MRI exams. Our post-processing module enabled an important gain on the 9 metrics, for both pipelines. The dual binary pipeline obtained better metrics on the liver than the multi-class one. However, for tumor, the multi-class AHUnet is slightly superior. Finally, the multi-class pipeline relying on the DynUnet (ablation study) is worst than the multi-class AHUnet for all metrics, showing the relevance of using a neural network specifically designed for anisotropic data. 

To assess if the differences between both proposed pipelines are statistically significant, we performed a signifance analysis, presented in Table \ref{tab:significance}. We first test the hypothesis that the obtained scores are drawn from a normal distribution (Shapiro's test). If the hypothesis is valid, one-sided student T-test are performed, else we opt for Wilcoxon's tests. Results show that the Liver metrics are significativelly better for the Dual binary pipeline than for the multiclass pipeline. On the contrary, the Tumor HD is significatively lower for the multi-class pipeline than for the binary one. Other metrics are not statistically different between both pipelines (Tumor ASD, Dice, HD and RMSE). 

Finally, the qualitative assessment of our lesion uncertainty module (Figure \ref{fig:uncertainty}, left) shows that FP lesions tend to be attributed to higher  uncertainty scores than TP lesions). The proposed score could thus be used to draw the user's attention to uncertain lesions, that are more likely to be FP. The proposed lesion uncertainty score is highly correlated with the lesion volume (Figure \ref{fig:uncertainty}, right), reaching a Spearman correlation coefficient of $-0.82$, which is intuitive as FP lesions are more frequent among small lesions. To complement this analysis, we provide two examples of the proposed lesion quantification in Supplementary Material.

\setlength{\tabcolsep}{5pt}
\begin{table}[]
\caption{Significance analysis between the two pipelines. B: binary pipeline; M: multi-class pipeline.}\label{tab:significance}
\centering
\begin{tabular}{c|l|ccccc}

& & \begin{tabular}[c]{@{}c@{}} Shapiro \\ p-value \end{tabular} & Hypothesis & Test & p-value & \begin{tabular}[c]{@{}c@{}} Significance \\ ($p<0.05$) \end{tabular} \\ \hline
\parbox[t]{2mm}{\multirow{4}{*}{\rotatebox[origin=c]{90}{Liver}}} & ASD $ \downarrow$ & $4.9\times 10^{-13}$ & B < M & Wilcoxon & $6.2\times10^{-4}$ & \checkmark \\
& Dice $\uparrow$  & $7.1\times 10^{-14}$& B > M & Wilcoxon & $7.1\times10^{-14}$ & \checkmark \\
&  HD $\downarrow$  &  $3.0\times 10^{-4}$ & B < M & Wilcoxon & $4.6\times10^{-2}$ & \checkmark \\
& SD $\uparrow$ & $1.9\times 10^{-9}$ & B > M & Wilcoxon & $1.1\times10^{-7}$ & \checkmark \\ \hline
\parbox[t]{2mm}{\multirow{5}{*}{\rotatebox[origin=c]{90}{Tumor}}} & ASD $\downarrow$ & $3.0\times 10^{-7}$ & M < B & Wilcoxon & $2.0\times10^{-1}$  & $\times$ \\
 &  Dice $\uparrow$  & $2.3\times 10^{-4}$ & M > B & Wilcoxon & $2.3\times10^{-1}$ & $\times$ \\
 & HD $\downarrow$ & $4.1\times 10^{-4}$ & M < B & Wilcoxon & $1.4\times10^{-3}$ & $\checkmark$\\
 & SD $\uparrow$ & $3.8\times10^{-1}$ & B > M & T-test & $7.6\times10^{-1}$ & $\times$ \\
 & RMSE $\downarrow$ & $2.7\times10^{-8}$ & B < M & Wilcoxon & $7.7\times10^{-1}$ & $\times$\\
\end{tabular}
\end{table}

\section{Conclusion}
In this work, we addressed the problem of segmenting the liver and tumors in CE-MRI. Using the AHUnet architecture as cornerstone, we propose two different pipelines, multi-class and dual binary, demonstrating a gain on tumor and liver segmentation respectively. Additionally, we propose to quantify the uncertainty of identified nodules, which can be used to improve the interpretability of the automated predictions. For the final ATLAS 2023 challenge submissions, models trained on the 5 folds are aggregated in an ensemble, for each pipeline separately. Both pipelines (binary and multiclass) are submitted as separate algorithms.

\section{Supplementary Material}

\begin{figure}
    \centering
    \includegraphics[width=\textwidth]{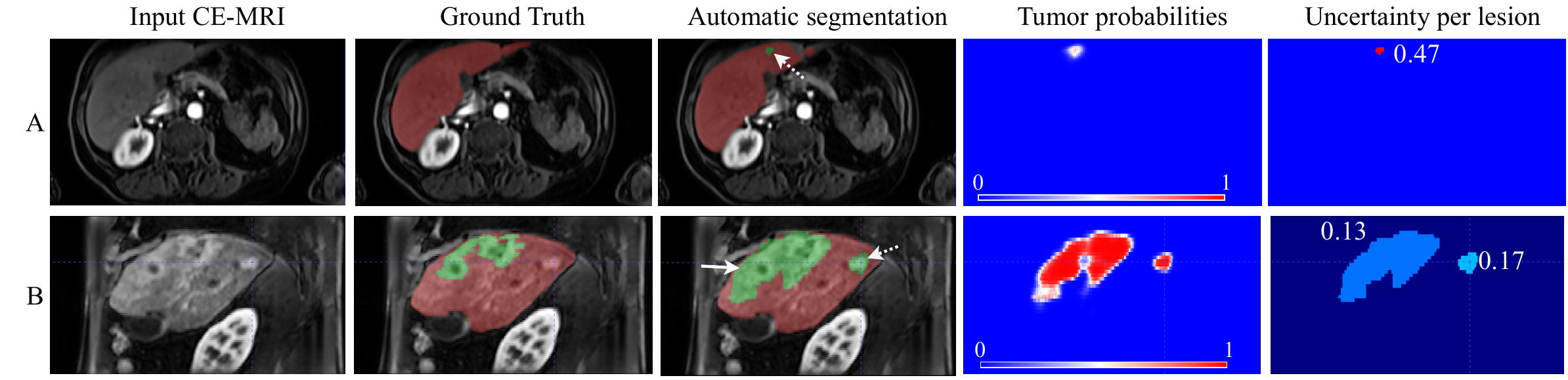}
    \caption{Case A: patient with a FP lesion (dashed arrow) associated with a high uncertainty value ($=0.47$); Case B: patient presenting one TP lesion (arrow) associated with a low lesion uncertainty (0.13) and a second FP lesion (dashed arrows) associated with a low uncertainty (0.17) in hypersignal on arterial phase. Complementary information (precontrast, portal and delayed CE-MRI T1 phases) are required to assess if this nodule is a CHC.}
    \label{fig:supp_unc}
\end{figure}


%
%

\bibliographystyle{splncs04}
\bibliography{biblio}

\end{document}